# De-orbiting Small Satellites Using Inflatables

**Aman Chandra**
*Space and Terrestrial Robotic Exploration Laboratory,*
*Department of Aerospace and Mechanical Engineering, University of Arizona*
**Jekanthan Thangavelautham**
*Space and Terrestrial Robotic Exploration Laboratory,*
*Department of Aerospace and Mechanical Engineering, University of Arizona*

## ABSTRACT

Small-satellites and CubeSats offer a low-cost pathway to access Low Earth Orbit at altitudes of 450 km and lower thanks to miniaturization and advancement in reliability of commercial electronics. However, at these low altitudes, atmospheric drag has a critical effect on the satellite resulting in natural deorbits within months. As these small systems further increase in reliability and radiation tolerance they will be able readily access higher orbits at altitudes of 700 km and higher, where atmospheric drag has little to no effect. This requires alternative technologies to either de-orbit these small spacecrafts at the end of life or move them to a safe parking orbit. Use of propulsion and de-orbit mechanisms have been proposed, however they require active control systems to be trigged. Other typical de-orbit mechanism relies on complex mechanisms with many moving parts.

In this work, we analyze the feasibility of using inflatable de-orbit devices that are triggered passively when a spacecraft is tumbling. Inflatables have already been proposed as hypersonic deccelerators that would carry large payload to the Martian surface. However, these systems are quite complex and need to withstand high-forces, temperature and enable survival of a critical payload. Furthermore, inflatables have been proposed as communication antennas and as structures using a class of sublimates that turn into gas under the vacuum of space. These inflatables system are relatively simple and does not require a specialized inflation system. Furthermore, these inflatable can be rigidized using UV curable resin that hardens the inflatable shell.

Inflatables offer the best mass to volume ratio and can be hardened to form solid shells. The proposed inflatable de-orbit device needs to perform several functions, including reducing tumbling followed by setting the dead spacecraft on a path towards orbit degradation. Tumbling reduction requires use of passive mechanisms to provide a suitable counter-torque to a tumbling spacecraft. This can be achieve using Solar Radiation Pressure (SRP). A large enough drag area needs to be created by the inflatable to gradually reduce the orbit of a small satellite and have it on a predictable deorbit trajectory. A large footprint produced by the de-orbit device simplifies ground based tracking.

We analyze the impact of deploying meter sized inflatable as de-orbit device. Furthermore, we analyze the total footprint of the de-orbit device on the spacecraft. Alternately, we also consider combined de-orbit device and inflatable communication antennas to determine its figure-of-merit. Our studies show potential feasibility of an inflatable deorbit device for CubeSats at higher altitude in Low Earth Orbit. However, certain unknowns are the subject of further study including long-term effects of UV and solar radiation on the inflatable membranes.

## 1. INTRODUCTION

Small satellites such as CubeSats are providing low cost access to space. Built to well-defined reference standards [1], they allow use of commercial off the shelf (COTS) components greatly reducing cost and development time. CubeSats are being used extensively in Low Earth Orbits (LEO) however in higher orbits, they are faced with limitations in COTS electronics not being rugged enough to tolerate radiation from trapped ionized particles in the Van Allen Belt. Recent advances in radiation hardened micro-electronics [2] have allowed CubeSat electronics to be used reliably in Geosynchronous Earth Orbits (GEO) and beyond. This opens the doors to CubeSats being widely used for Earth-orbiting space missions. The launch of the MarCO CubeSats in 2018 is enabling evaluation of CubeSat subsystems and electronics in deep space [3]

Increased standardization of electronics has made space technology readily accessible at an unprecedented scale. There has been an exponential increase in the number of CubeSat missions being sent into space since 2010. Their popularity has triggered concerns over orbit usage especially in LEO. Studies have shown that increasing launch opportunities have led to the placement of 60-70 new objects per year into LEO [4]. Extrapolation of this data suggests that the LEO environment may see a tipping point called the "Kessler Effect" in a few decades from now. The Kessler Effect would result in saturation of objects in LEO leading to uncontrolled phenomenon such as cascading collisions and communication blockages. It is necessary, therefore, to provide traffic management and service Low Earth Orbits to remove orbiting debris. Fig. 1 is a visual illustration of LEO congestion.

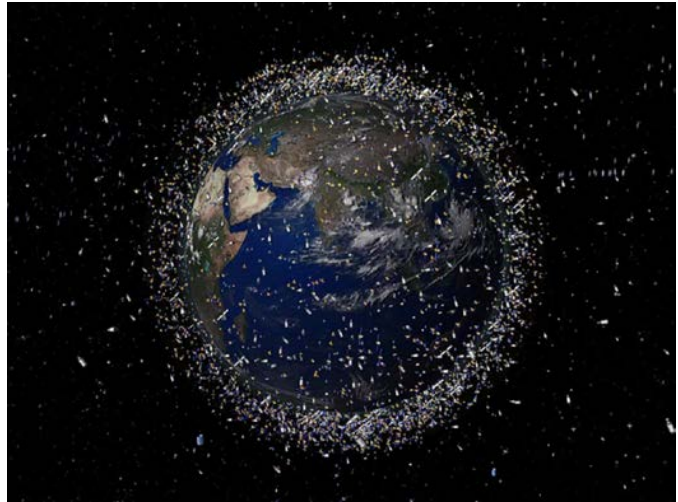

Fig. 1. Visual illustration of the "Kessler Effect"

Several strategies have been researched for the servicing of orbits. These include Active Debris Removal (ADR) and On-Orbit Satellite Servicing (OSS) [5]. ADR requires placing an end effector such as a tether or net in the desired orbit dedicated solely to capture and harness debris. This method does not prove very effective for debris disposal. OSS includes methods of placing robotic components on existing satellites to facilitate orbital debris collection and disposal. Both these methods face the common hurdle of increasing the complexity of mission design, increased cost and logistics. Constraints imposed by limited volume and mass available on CubeSats makes them unfeasible.

A far more efficient strategy is to design an orbital system capable of de-orbiting and 'self-servicing' at the end of the mission. CubeSats in low earth altitudes encounter aerodynamic drag leading to orbital decay over time resulting in burn up and disposal in the Earth's atmosphere. For altitudes above 500 km, however, atmospheric density reduces significantly. This could lead a CubeSat to remain in orbit beyond 25 years, the current mandated limit for orbit lifetimes. Increasing encountered drag for small satellites is key to reducing their orbital lifetimes. Thrusters have been used to provide propulsive deceleration. In the case of CubeSats, however, including a propulsion system and allocating a considerable volume for propellant may not always be practical.

This leads us to deployable structures to achieve required aerodynamic drag. Among deployable structures, inflatable gossamers offer the highest packing efficiency. Such structures are ultra-light and offer very high deployed surface to mass ratios. Additionally, they can be scaled to sizes in the order of meters. These attributes make it possible to achieve low ballistic coefficients with inflatables. Table 1 shows a comparison of various deployable technologies for braking.

Table 1. Comparison of deployable braking technologies

| Braking technology | Mass/ unit surface area (kg) | Packing ratio | Drag coefficient | Ballistic coefficient (Pa) |
|---|---|---|---|---|
| Deployable panels | 0.5 – 0.7 | 2:1 – 3:1 | 2 - 4 | 150 - 200 |
| Linkage systems | 0.7 - 1 | 5:1 – 8:1 | 3 - 4 | 20 - 100 |
| Inflatable gossamers | 0.06 – 0.2 | 15:1 – 20:1 | 2 - 4 | 5 - 20 |

In this paper, we look into feasible design configurations for a CubeSat based inflatable aerodynamic drag/braking device. A design strategy is evolved to produce concept braking structures compatible with the CubeSat form factor. Deceleration requirements narrow down the structural design space for the inflatables. Structural response studies are conducted to understand important design traits with respect to aero-braking performance. We then compute the drag co-efficient and ballistic co-efficient of the proposed design to access expected braking performance. A discretized structural model is then presented to analyze the effect of scale on aero-braking and structural performance for different geometries. In the following sections we present related work, followed by methodology, results, discussion, conclusions and future work.

## 2. RELATED WORK

Deployable systems have been studied for atmospheric re-entry. However, work dedicated to aero-braking for on orbit servicing has received limited attention. Inflatable structures have seen development since the 1950's when NASA launched their ECHO satellite balloon program [6]. Inflatable technology received considerable attention for structural applications varying from gossamer sails, antennas, landing airbags and solar panels [7]. Ruggedized inflatables made of thermal fabrics started being researched for the challenging thermos-structural conditions during atmospheric re-entry. The first inflatable re-entry test was carried out in the year 2000 as a demonstration of inflatable re-entry and descent technology (IRDT) [8]. The structure consisted of an inflatable flexible heat shield and a parachute landing system. The experiment demonstrated significant improvement in payload to mass ratios due to highly efficient packing and low-weight of the inflatable. Structural and thermal performance was observed to be enough to survive atmospheric-entry into Earth.

The success of the IRDT established inflatable technology as a robust and low-cost alternative to existing re-entry technologies. A second successful experiment was seen when NASA launched the inflatable re-entry vehicle experiment (IRVE) in 2006. The IRVE is a 3-meter diameter, 60° half angle sphere cone consisting of an inflatable aero-shell structure. The purpose of the experiment was to validate aero-shell performance for atmospheric re-entry [9].

Successful tests with inflatables for hypersonic re-entry established inflatables as a promising, ultra-lightweight rugged solution. This led to research into smaller scale inflatables on small satellites including CubeSats. For a braking or de-orbit device, the encountered thermo-structural loads are an order of magnitude lower than atmospheric re-entry. In our earlier, we analyzed the feasibility of atmospheric entry into Mars using a drag-device that needs to withstand high-temperatures [20]. Such as small system would then deploy small robots such as the SphereX platform onto the surface of Mars [21-22]. Here as we are just analyzing drag device, this simplifies the design and allows compact and lightweight gossamers to be used. Andrews Space has developed a prototype that has undergone ground based tests as an inflatable nanosat de-orbit and recovery system for CubeSat payloads [10]. Fig. 3 shows an illustration of their design. Italian concept IRENE [11] is undergoing tests with a spherical cone designs but is intended for much larger payloads.

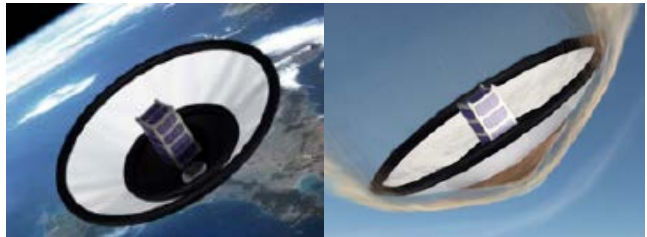

Fig. 3. Proposed d2U CubeSat DRS mechanism [10]

Carandente et al [5] present concepts for aero-braking structures from a de-orbit and re-entry point of view. The concepts presented are ones that offer substantial reduction of ballistic coefficient of these structures. The authors point out that a lower ballistic coefficient leads to a reduction in peak mechanical and thermal loads experienced by the structure while achieving higher deceleration rates. This highlights the potential of using large inflatable gossamers that can be packaged into very small volumes for nano-satellite payloads.

Among large scale gossamer structures, Global Aerospace Corporation proposed the Gossamer Orbit Lowering Device (GOLD) to de-orbit spent stages and old or derelict satellites [12]. The concept consists of deploying a large inflatable sphere several meters in diameter that offers exceptionally low ballistic coefficients in Lower Earth Orbits. Fig. 4 shows the concept. Gossamer sails made of Kapton have also been studied in considerable detail [13]. While sails potentially offer more efficient packing ratios than inflatables, their structural reliability for aerobraking is not clearly established. While encountering loads due to atmospheric drag, a pneumatic pressure system has been used to provide necessary resistive stiffness. In the case of sails, additional structural re-enforcement is needed which has reduces packing efficiency and increases deployment complexity.

Pneumatic inflatables have shown robust structural behavior while maintaining ease of scaling into large sizes. The focus of our research is on inflatable structures. Pneumatic inflatable require a gas source. This can be in the form of a compressed gas or gas producing chemical reaction. Inflatables using solid state sublimates as gas sources have shown promising results for Low Earth Orbit operation [14]–[16]. Our present work focuses on structural design using a sublimate gas source.

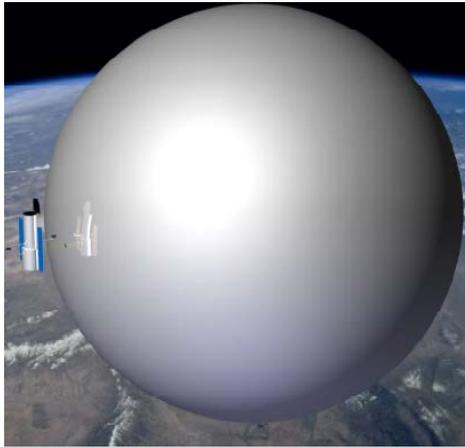

Fig. 4. GOLD system for de-orbiting a large observatory [12]

### 3. METHODOLOGY

An effective design strategy for inflatable de-orbit devices would require thorough understanding of structural loads encountered by such devices. The loads would depend on the geometry and stiffness of the structure. Our methodology begins with characterizing expected loads that would need to be countered by the inflatable. Since these loads are a function of the structures geometry, this serves to provide us with bounds on geometrical constraints. Geometric and structural design bounds thus lead us to evaluate several candidate designs for the de-orbiting device. A discrete finite element method is used for an approximate analysis of structural performance. This helps us understand the extent to which these structures can be scaled up and the corresponding effect on their aero-braking performance.

**3.1 Environmental Loads and Geometric Requirements**

De-orbit performance is proportional to the drag force experienced by the device. As the spacecraft altitude increases, the surrounding atmosphere continues to rarify thereby reducing drag. The effect of height on atmospheric drag is described by (1)

$$\rho \approx \rho_o e^{-\Delta h / h_o(h)} \qquad (1)$$

The atmospheric density $\rho$ at a given altitude and $\rho_o$ at a second altitude with difference in height of $\Delta h$ are related exponentially as shown. $h_o(h)$ termed as scale height is a function of altitude. We begin by studying the nature of forces encountered towards two major applications. For a circular orbit at altitude $H$ above the Earth, the average change in acceleration due to drag is as shown in (2)

$$\Delta a_{rev} = -\frac{2\pi\rho a^2}{b_c} \qquad (2)$$

To enable atmospheric burn up at 100 km altitude, requires the ballistic coefficient be a function of altitude and can be written as:

$$b_c = \frac{(R_e + H)^2 (2R_e + H + 10^5)(H - 10^5)}{2\pi\mu\rho(R_e + 10^5)^2} \qquad (3)$$

### 3.2 Structural Design Methodology

Our design of the aero-braking structures is based on competing structural requirements. The structure must be strong enough to withstand the braking force but must also be able to efficiently transfer loads to prevent stress localization and consequent failures. Another major requirement is thermal endurance. Dorsey and Mikulas [17] present their work on fundamental design and sizing equations developed for aerobraking structures. For support structures with no curvature, lift to drag ratios can be assumed to be less than 0.5 [18]. Traditionally, two separate strategies have been developed. One is the design on the support structure and the other is that of the shield that offers resistance and drag. The support truss is designed to transfer aerodynamic drag from the braking shield structure. Fig. 5 shows a schematic diagram of this basic configuration:

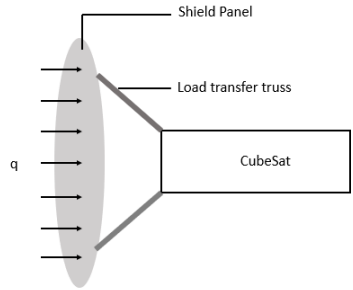

Fig. 5. Structural design configuration.

The pressure distribution on the surface of the shield can be assumed to be uniform in nature and is calculated from the aerobrake inertia force due to a constant rate of deceleration. The pressure $p$ is given as shown in (4) as:

$$p = \frac{M_s a}{g A_{ab}} \qquad (4)$$

Here $M_s$ is the mass of the spacecraft attached to the braking system, $a$ is the deceleration rate, $g$ is a constant defined by Newton's law as Force = (mass × acceleration). $A_{ab}$ is the area of the aerobraking structure. We extend their methodology to include design of inflatable membrane structural units. Based upon structural function, the aero-braking device consists of a shield and support structure. The fundamental structural sizing equation is a shown below.

$$w_{max} = \alpha \frac{qA^4}{D_{HP}} \qquad (5)$$

Here $w_{max}$ represents a bound on maximum mass of the structure for achieving a bending stiffness $D_{HP}$ for encountered drag force $q$ over area $A$. Based on sizing requirements, we propose two spherical cone based structural concepts as shown in Fig. 6 and 7 below:

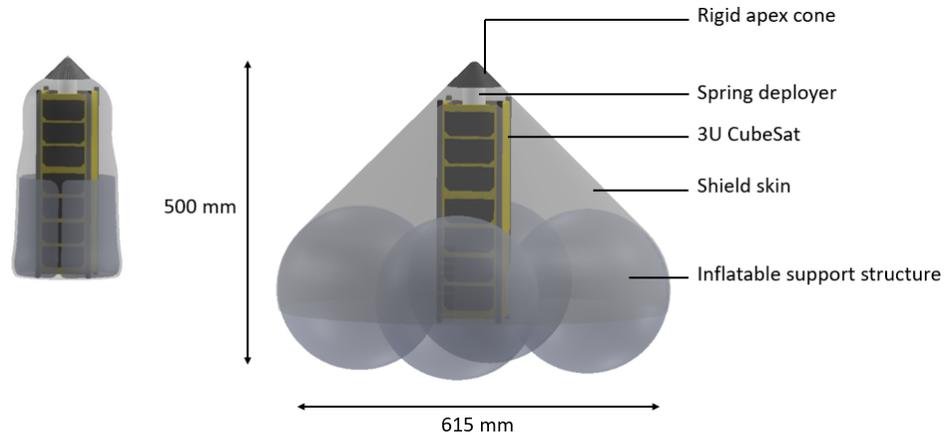

Fig. 6. Concept 1 with spherical inflatable aeroshell.

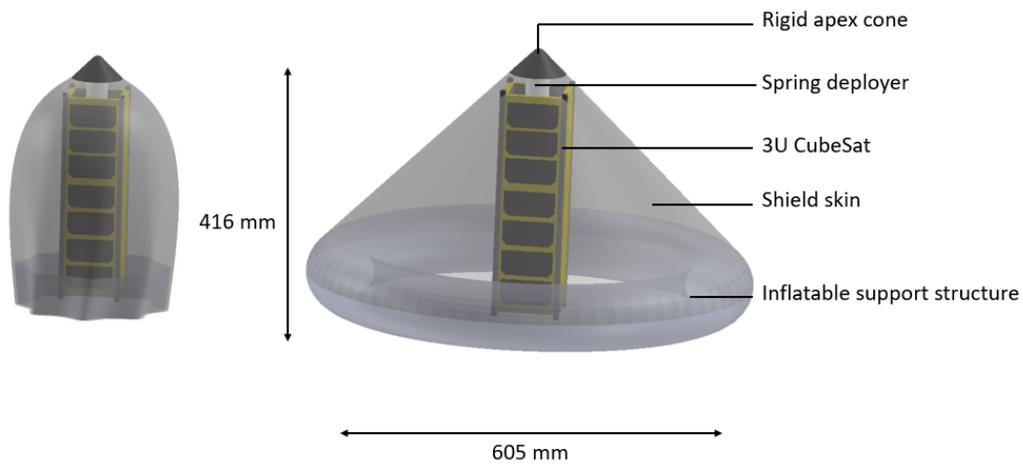

Fig. 7. Concept 2 with toroidal inflatable aero-shell.

Our concepts consist of inflatable structural members providing load transfer from a membrane skin. The apex cone acts as rigid support on the other end. The rigidity of the apex is to withstand thermo-structural load concentration at the apex region. Concept 1 uses an assembly of inflatable spheres as support while in Concept 2 that has been replaced by an inflatable toroid.

**3.3 Design Performance Analysis**

The design's performance is analyzed to understand the structural response and aerobraking performance. The inflatable membrane's structural response was simulated using commercial finite element software LS-Dyna. Stress and deformations were obtained for each individual element. To assess the performance of the inflatable membranes, a separate strategy was adopted. Fig. 8 shows a discretized simplification of an inflatable membrane. The inflatable element has been replaced by a 3-dimensional truss element termed as a voxel. This replaced the spherical inflatable with a tetrahedral structure. While this is not an accurate representation of membrane dynamics, it serves to understand the effect of geometry and scale on the structure's response.

Aero-braking or de-orbit performance of the inflatable is characterized by estimating its drag co-efficient. The following equation is used to assess the drag co-efficient based on the structure's geometry:

$$C_D = \alpha \left( \frac{4}{A} \int_S \cos^3 \phi \, dS \right) + \beta \left( 2 + \frac{1}{A} \int_S \cos^2 \phi \, dS \right) + 2\gamma \tag{6}$$

$\alpha$, $\beta$ and $\gamma$ represent relative fractions contributing to specular, diffuse and absorptive surfaces on the structure. Hence, the sum $\alpha+\beta+\gamma$ equals 1. $\Phi$ represents the angle between the structure's surface normal and velocity vector. Computed drag coefficient values were used to calculated obtained deceleration using (7).

$$\Delta P_{rev} = -6\pi^2 \left( C_D A / m \right) \rho a^2 / V \tag{7}$$

Here $\Delta P$ represents a change in orbital period for a circular orbit characterized by the CubeSats velocity V and mass m. The calculated loads are compared with expected structural behavior to understand their ability to maintain structural integrity.

### 3.4 Pathway Towards Large Scale Structures

Conducting full scale structural simulations on large inflatable assemblies is time intensive and not computationally efficient. An alternate strategy is to perform reduced order simulations using discretized approximation. As shown in Fig. 9 , the discretized structure consists of a truss assembly also known as a voxel [19]. We use this method to understand general structural effects on varying geometry and size of inflatable structures.

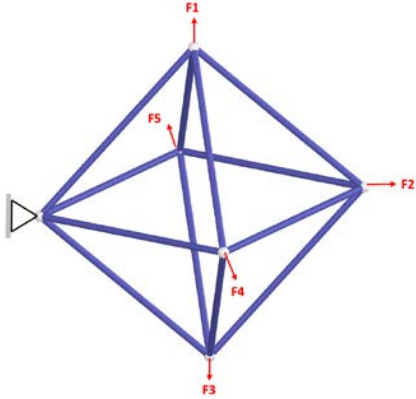

Fig. 8. Discretized voxel approximation.

6-DoF (Degree of Freedom) element stiffness matrices were computed for each element in the voxel as shown in (8).

$$K^e = R^{e^T} k^e \begin{bmatrix} 1 & -1 \\ -1 & 1 \end{bmatrix} R^e \tag{8}$$

Where $k^e$ represents the stiffness of each element and $R^e$ represents a transformation matrix given by:

$$R^e = \frac{1}{l^e} \begin{bmatrix} x_n^e & y_n^e & z_n^e & 0 & 0 & 0 \\ 0 & 0 & 0 & x_n^e & y_n^e & z_n^e \end{bmatrix}$$

Here $l$ represents the length of each element and $x$, $y$ and $z$ represent change in $x$ and $y$ and $z$ coordinated between adjacent elements in the voxel. The elemental stiffness matrices are scattered and summed to form the global stiffness matrix as follows:

$$K = \sum_e L^{eT} K^e L^e$$

$L^e$ represents a binary matrix to scatter stiffness values over the length of the displacement vector. Nodal displacements are then found using the following equation:

$$d = K^{-1} F \qquad (9)$$

## 4. RESULTS AND DISCUSSION

This section discusses results obtained from our analysis. We begin by simulating the final obtained shape of the drag devices as shown in Fig 10 and 11. This is referred to as the free-form structural response. The final deployed shape obtained from these simulations then becomes a basis for calculations of drag and ballistic coefficients that are used to characterize aero-braking performance of the inflatable.

### 4.1 Free-form Structural Response of Inflatable Concepts

Free form analysis of deflated membranes was conducted to understand their inflated geometries. The final shape of the aeroshell was determined from best fit obtained geometries. Fig. 10 shows evolved geometry of Concept 1 over time. The spherical supports attached to the skin are observed to extend outward and move outward away from the CubeSat's walls as shown.

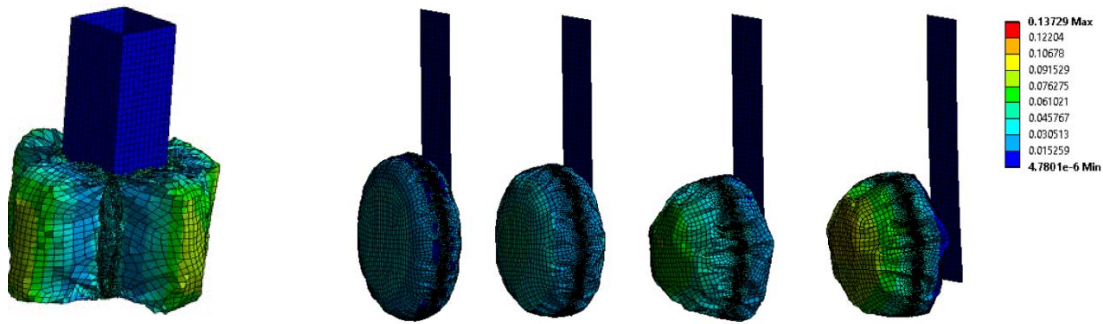

Fig. 10. Deformation plot for Concept 1.

Fig. 10 shows evolved geometry of Concept 2 over time. Deformation observed in this case is much lower as compared to Concept 1. This is due to a uniform distribution of stress across a larger surface area offered by Concept 2. As can be seen in Fig. 12, this also translates into much slower stress in the case of Concept 2.

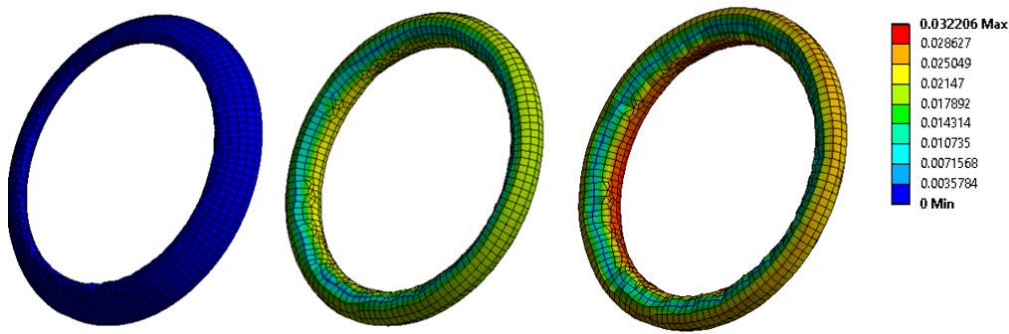

Fig. 11. Deformation plot for Concept 2.

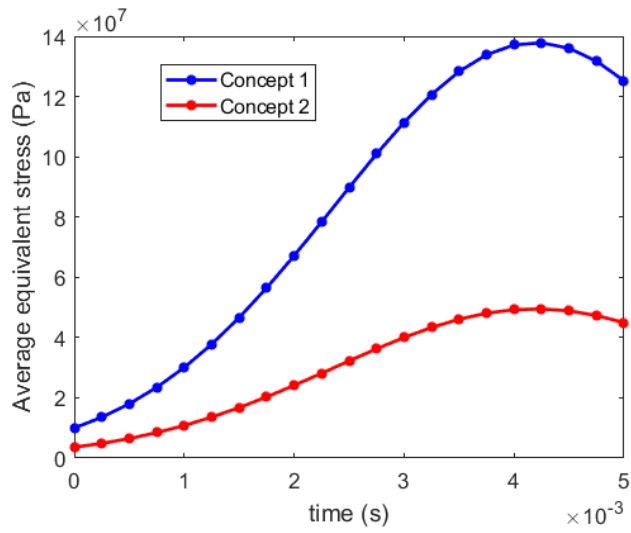

Fig. 12. Stress comparison plots for Concept 1 and 2.

It can be observed that the overall stresses developed due to aero-braking loads are well above critical or buckling stresses in both inflatable designs.

**4.2 Aero-braking Performance**

Table 2 shows estimated drag coefficients, ballistic coefficients and estimated orbit decay for a 3U CubeSat using both inflatable concepts with an estimated total mass of 4 kg.

Table 2. Comparison of braking concepts

| Design Concept | Mass (kg) | Surface Area (m2) | Drag Co-efficient | Ballistic Co-efficient (kg/m2) |
| --- | --- | --- | --- | --- |
| 3U CubeSat | 3.5 | 0.01 | 2 | 175 |
| Concept 1 | 4 | 0.248 | 2.667 | 6.05 |
| Concept 2 | 4 | 0.346 | 2.7 | 4.28 |

The above table shows a dramatic decrease in ballistic co-efficient upon adding the inflatable structures onto the 3U CubeSat. This is due to much larger surface areas at very low additional mass. A larger drag coefficient is possible in thanks to an optimized spherical cone geometry. Based on calculated co-efficient values, we go on to calculate estimated de-orbit lifetimes for each case. Table 3 shows estimated orbit decay times from various altitudes of a circular orbit for both design concepts.

Table 3. Comparison of expected de-orbit times

| Altitude (km) | Disposal Life-time (years) | | |
| --- | --- | --- | --- |
| | 3U CubeSat | Concept 1 | Concept 2 |
| 400 | 1.2 | 0.05 | 0.045 |
| 500 | 6.3 | 0.3 | 0.28 |
| 600 | 23.5 | 1.58 | 1.5 |
| 700 | >25 | 6.4 | 6.2 |
| 800 | >25 | 18.5 | 17.8 |

The reduction in de-orbit lifetimes is in agreement with much lower ballistic co-efficient. Concept 2 is found to have greater braking and structural performance in comparison with Concept 1.

## 5. CONCLUSION AND FUTURE WORK

Our work demonstrates a structural design strategy for the design and development of inflatable gossamer aero-braking structures for de-orbiting small satellites. Based on structural design requirements, we proposed two spherical cone designs. Our concept consists of a membrane skin supported on a rigid-apex on one end and an inflatable support structure on the other. Of the inflatable support structures, two separate concepts were put forth. The first was inflatable spheres and the second was an inflatable toroid. Structural analysis reveals much more uniformly distributed loads on the toroidal design leading to a reduction in net stresses.

When it comes to aero-braking performance, the toroidal design was observed to offer greater encountered surface area for the same mass leading to very low ballistic co-efficient. Our analysis shows clear advantages of a toroidal inflatable supported gossamer aero-braking structure. The methodology presented leads to the successful conceptual design of a nano-satellite based de-orbiting device. We have gone on to analyze scaled up assembles up to 5 meters in diameter using discrete finite elements. This has helped identify structural traits necessary to ensure scaling up of inflatable de-orbit device for usage on large satellites and telescopes.

Further work includes incorporating high fidelity atmospheric models to estimate drag coefficients with greater accuracy. This would be used to further refine the structural design of the proposed concepts. The structural model will need modifications to incorporate thermal loads and thermal stress concentrations on the inflatable device. This will be followed by testing in a laboratory followed by testing under a relevant environment.